\documentclass[a4paper,12pt]{article}
\usepackage{amsmath,amsfonts,amssymb,cite}
\usepackage[dvips]{graphicx}

\renewcommand{\Im}{\text{Im\,}}
\renewcommand{\L}{\textbf{L}}
\newcommand{\als}{\alpha_s}
\newcommand{\qqb}{\langle\bar{q}q\rangle}

\newcommand{\mqq}{\langle m_q\bar{q}q\rangle}

\newcommand{\gmv}{\left\langle\frac{\als}{\pi}\left(G_{\mu\nu}^a\right) ^2\right\rangle}
\newcommand{\lb}{\left(}
\newcommand{\rb}{\right)}
\newcommand{\lsb}{\left[}
\newcommand{\rsb}{\right]}

\begin{document}

\begin{center}
{\Large\bf
%Slope of radial Regge trajectories from borelized spectral sum rules
The large-$N_c$ limit of borelized spectral sum rules and the slope of radial Regge trajectories}
\end{center}
\bigskip
\begin{center}
{\large S. S. Afonin
%\footnote{Email: \texttt{s.afonin@spbu.ru}.}
and T. D. Solomko}
\end{center}

\begin{center}
{\it Saint Petersburg State University, 7/9 Universitetskaya nab.,
St.Petersburg, 199034, Russia}
\end{center}
%\bigskip

\begin{abstract}
%The particle phenomenology and various hadron models suggest that
%radially excited light mesons lie on approximately linear Regge-like
%trajectories with nearly universal slope. The value of slope is an
%important quantity in non-perturbative QCD and gives rise to the
%scale of meson masses in the light quark sector.
We put forward a
new phenomenological method for calculating the slope of radial
trajectories from values of ground states and vacuum condensates.
The method is based on a large-$N_c$ extension of
borelized spectral sum rules. The approach is applied to the light
non-strange vector, axial, and scalar mesons. The extracted values
of slopes proved to be approximately universal and are in the interval
$1.4\pm0.1$~GeV$^2$. As a by-product, the given method leads to
prediction of the second radial trajectory with ground state mass
lying near 0.6~GeV.
\end{abstract}

\section{Introduction}

It is widely believed that confinement in QCD leads to approximately linear
radial Regge trajectories (see, e.g.,~\cite{phen,phen2}). The linearity has a natural
explanation within various hadron string models~\cite{string}. The most important
quantity in this picture is the slope of trajectories. The slope is expected
to be nearly universal as arising from flavor-independent non-perturbative
gluodynamics which thereby sets a mass scale for light hadrons. In view of
absence of analytical description for hadron mass generation, it is
interesting to construct phenomenological methods that would allow to
estimate the value of slope and check its universality basing on some inputs from QCD.
Among the phenomenological approaches to the hadron spectroscopy, the method
of spectral sum rules~\cite{svz} is perhaps the most related with QCD.
In many cases, it permits to calculate reliably the masses of ground
states on the radial trajectories. We propose an extension of this approach which
allows to estimate the slope using essentially the same technic.

The method of QCD sum rules was originally introduced by
Shifman, Vainshtein and Zakharov (SVZ)~\cite{svz} and it
turned out to be extremely fruitful in the hadron spectroscopy~\cite{rry,colangelo}.
The idea of this approach is based on the assumption that a quark-antiquark
pair being injected into the strong QCD vacuum does not perturb appreciably
the vacuum structure. This allows to parametrize the unknown properties of
non-perturbative vacuum by some universal phenomenological
characteristics called vacuum condensates. According to this approach,
hadrons with different
quantum numbers have different masses and other static characteristics
because their currents react differently
with the vacuum medium. A manifestation of this difference
are different coefficients in the Operator Product Expansion (OPE)
of correlators of the corresponding quark currents which
can be calculated from QCD. Assuming the
existence of resonance in some energy range, one is able to
calculate its static characteristics via the dispersion relations and the
OPE, with higher radial excitations being regarded as a part of perturbative continuum.
In the case of the light hadrons, a borelized version of sum rules is usually
exploited as the Borel transform effectively singles out the ground state
suppressing contributions from the rest of spectrum and simultaneously improving the
convergence of OPE~\cite{svz}.

The SVZ sum rules does not allow to calculate the full decay width
since hadrons are considered as infinitely narrow states. A
typical accuracy of the method is thus of the order of 10 - 20\%.
On the other hand, the narrow-width approximation for mesons has a
solid theoretical basis --- the large-$N_c$ (planar) limit in
QCD~\cite{hoof,witten}. In this limit, the one-hadron states
saturate completely the two-point  correlation functions of hadron
currents $j$,
\begin{equation}
\label{1}
\left\langle j(q)j(-q)\right\rangle=\sum_n\frac{F_n^2}{q^2-M_n^2}.
\end{equation}
The large-$N_c$ scaling of quantities is:
$M_n=\mathcal{O}(1)$ for masses,
$F_n^2=\langle0|j|n\rangle^2=\mathcal{O}(N_c)$ for residues,
$\Gamma=\mathcal{O}(1/N_c)$ for decay width. The sum in~\eqref{1}
must contain an infinite number of terms in order to reproduce
the logarithmic behavior of the correlator at large $q^2$
following from the asymptotic freedom~\cite{witten}.
Assuming some ansatz for
the radial mass spectrum, the expression~\eqref{1} can be summed up,
expanded at large $Q^2=-q^2$ and compared with the corresponding
OPE in QCD. One obtains a set of sum rules ---
each sum rule represents an equation corresponding to
some $k$ in the expansion $1/Q^{2k}$, $k=0,1,2,\dots$ of both
sides in~\eqref{1}.
Such planar sum rules were considered many
times in the case of the linear Regge ansatz for radial
spectrum motivated by the phenomenology~\cite{sr,AS},
\begin{equation}
\label{2}
m_n^2 = a n + m_0^2, \qquad n=0,1,2,\dots,
\end{equation}
sometimes with certain non-linear corrections to this spectrum.

This planar approach to QCD sum rules has a certain shortcoming: The sum rules
corresponding to different $k$ are treated on equal footing while
the accuracy of sum rules deteriorates rapidly with increasing $k$
because of a bad convergence.
In the given paper, we propose a novel treatment of the planar QCD
sum rules. The idea is to apply the Borel transform to the infinite sum
in the r.h.s. of~\eqref{1} and derive an expression for the slope $a$
in~\eqref{2} in the same way as one finds an expression for a mass
of ground state in the classical SVZ sum rules. In other words,
we propose to consider borelized planar sum rules and
analyze them following a well elaborated technics. We will consider $m_0$
in~\eqref{2} as the mass of ground state. The value of $m_0$ will be
regarded as being known from the old SVZ sum rules or from experimental data.
In the first case, the standard calculation of $m_0$ within SVZ sum rules
represents just the first step to finding the whole radial spectrum.
We construct thus an extension of SVZ sum rules which allows to obtain
the radial spectrum using essentially the same number of input parameters.
The main output of the analysis is the slope $a$.
In essence, we propose a new way for calculating this quantity
with the help of a well developed method. We will apply this approach
to the light non-strange vector, axial and scalar mesons. In the latter case, our
approach leads to a unexpected result related with appearance
of the second scalar trajectory beginning with a rather light state
having mass near 600~MeV.

The paper is organized as follows. In Section~2, we formulate our approach.
Its application to the vector, axial, and scalar channel is considered in Section~3.
The conclusion and various remarks are given in Section~4.

\section{SVZ sum rules for the slope of radial trajectories}

\subsection{The vector case}

The case of vector $\rho$-mesons is canonical for the SVZ sum
rules method~\cite{svz}. Let us apply this method to the radial
spectrum~\eqref{2} in the large-$N_c$ limit.

The basic theoretical object is the two-point vector correlator
$\Pi(Q^2)$ in Euclidean space defined by
\begin{equation}
\label{3}
\Pi_{\mu\nu}=(q_\mu q_\nu-g_{\mu\nu}q^2)\Pi(Q^2),
\end{equation}
where $\Pi_{\mu\nu}$ represents the T-product of two vector
currents interpolating the neutral $\rho^0$-meson,
\begin{equation}
\label{4}
\Pi_{\mu\nu}=i\int d^4x\, e^{ipx}\left\langle0|\text{T}\left\{j_\mu(x),j_\nu(0)\right\}\right|0\rangle,
\end{equation}
$$j_\mu=\frac12(\bar{u}\gamma_\mu u-\bar{d}\gamma_\mu d).$$
Here $u$ and $d$ are quark fields, and $q$ is the photon spacelike
momentum, $q^2=-Q^2$. The OPE for $\Pi(Q^2)$ reads~\cite{svz},
\begin{equation}
\label{5}
\Pi(Q^2)=\frac{1}{8\pi^2}\left(1+\frac{\als}{\pi}\right)\ln\frac{\mu^2}{Q^2}
+ \frac{\mqq}{Q^4} +\frac{1}{24 Q^4}\gmv
-\frac{14}{9}\frac{\pi\als}{Q^6}\qqb^2,
\end{equation}
where $q$ stands for $u$ or $d$ quark, the coefficient in front of
the last term is given in the large-$N_c$ limit (it differs by the
common factor $(N_c^2-1)/N_c^2$ from the corresponding coefficient
in Refs.~\cite{svz,rry}), and further $\mathcal{O}(Q^{-8})$ terms
are neglected. Applying the Borel transform,
\begin{equation}
\label{6}
L_M\Pi(Q^2)=\lim_{\substack{Q^2,n\rightarrow\infty\\Q^2/n=M^2}}\frac{1}{(n-1)!}(Q^2)^n\left(-\frac{d}{dQ^2}\right)^n\Pi(Q^2),
\end{equation}
to the OPE~\eqref{5} we get
\begin{equation}
\label{7}
L_M\Pi(Q^2)=\frac{1}{8\pi^2}\left(1+\frac{\als}{\pi}\right) + \frac{\mqq}{M^4} +\frac{1}{24 M^4}\gmv -\frac{7}{9}\frac{\pi\als}{M^6}\qqb^2.
\end{equation}

The vector correlator $\Pi(Q^2)$ satisfies a dispersion relation
with one subtraction,
\begin{equation}
\label{8}
\Pi(q^2)=\frac{1}{\pi}\int_{4m_q^2}^\infty ds \frac{\Im\Pi(s)}{s-q^2 +i\varepsilon}+\Pi(0).
\end{equation}
In the large-$N_c$ limit, the mesons are infinitely narrow and
saturate completely the two-point correlators~\cite{witten},
\begin{equation}
\label{9}
\Pi(q^2)=\sum_{n}\frac{F^2_n}{q^2-m_n^2+i\varepsilon},
\end{equation}
where the residues are defined by $\langle 0|j_\mu|V_n\rangle=F_nm_n\varepsilon_\mu$.
The experimental information on the electromagnetic decay constants $F_n$
of radially excited light vector and axial mesons is poor. The quark-hadron
duality requires that $F_n$ must be constant for the exactly linear spectrum
(or decrease at least exponentially with $n$)~\cite{sr}. For the sake of simplicity
(a minimal number of inputs), we will assume that $F_n$ represent just a universal
constant in the large-$N_c$ limit, $F_n=F$. The given assumption will lead to rather
reasonable numerical predictions and this is enough for the zero-width approximation.
The imaginary part of $\Pi(Q^2)$ takes then a simple form
\begin{equation}
\label{10}
\Im\Pi(q^2)=\sum_{n}\pi F^2 \delta(q^2-m_n^2).
\end{equation}
The Borel transform of~\eqref{8} in this case is~\cite{svz}:
\begin{equation}
\label{11}
L_M \Pi(Q^2) = \frac{1}{\pi M^2}\int_{0}^\infty e^{-s/M^2}\Im \Pi(s)ds
= \frac{F^2}{M^2}\sum_{n}e^{-m_n^2/M^2},
\end{equation}
where we neglected the $\mathcal{O}(m_q^2)$ contribution.
Substituting the linear spectrum~\eqref{2} and summing up we
obtain
\begin{equation}
\label{12}
L_M \Pi(Q^2) = \frac{F^2}{M^2}\frac{e^{-m_0^2/M^2}}{1-e^{-a/M^2}}.
\end{equation}

The first sum rule arises from equating the relations~\eqref{5}
and~\eqref{12},
\begin{equation}
\label{13}
\frac{F^2  e^{-m_0^2/M^2}}{1-e^{-a/M^2}} =
\frac{M^2}{8\pi^2} \lsb
    1 + \frac{\als}{\pi} +
    \frac{8\pi^2}{M^4}\mqq +
    \frac{\pi^2}{3 M^4}\gmv -
    \frac{56}{9}\frac{\pi^3 \als}{M^6}\qqb^2
\rsb.
\end{equation}
Following the prescriptions of SVZ method~\cite{svz}, we can get
the second sum rule by taking derivative of Eq.~\eqref{13} with
respect to $1/M^2$. The meson mass appears directly in the
fraction $-\frac{d(13)}{d(1/M^2)}/(13)$. The given "combined" sum
rule has the form
\begin{equation}
\label{14}
m_0^2=M^2\frac{h_0 - \frac{h_2}{M^4} - \frac{2h_3}{M^6}}{h_0 + \frac{h_1}{M^2} + \frac{h_2}{M^4} + \frac{h_3}{M^6}}
-\frac{a}{e^{a/M^2}-1},
\end{equation}
where the condensate terms $h_i$ are presented in Table~1. In the
same table, we display the corresponding terms $h_i$ for
the axial and scalar cases~\cite{svz,rry}.

\begin{table}[ht]
\begin{center}
\caption{\small The condensate contributions $h_i$ in different meson channels.}
$\begin{array}{|c|c|c|c|c|}
 \hline
 \text{Mesons} & h_0 & h_1 & h_2 & h_3 \\
 \hline
 \rho & 1 + \frac{\als}{\pi} & 0 & 8\pi^2\mqq + \frac{\pi^2}{3}\gmv & -\frac{56}{9}\pi^3\als\qqb^2 \\
 \hline
 a_1 & 1 + \frac{\als}{\pi} & -8\pi^2 f_{\pi}^2 & -8\pi^2\mqq + \frac{\pi^2}{3}\gmv & \frac{88}{9}\pi^3\als\qqb^2 \\
 \hline
 f_0 & 1 + \frac{11}{3}\frac{\als}{\pi} & 0 & 8\pi^2\mqq + \frac{\pi^2}{3}\gmv & \frac{176}{9}\pi^3\als\qqb^2 \\
 \hline
\end{array}$
\end{center}
\end{table}

The first term in the r.h.s. of Eq.~\eqref{14} corresponds to the
limit $s_0\rightarrow\infty$ in the canonical expressions for the
meson masses in the SVZ sum rules~\cite{rry}. The energy cutoff
$s_0$ is infinite in our case as we take into account an infinite
number of radial excitations. The second term is new and reflects
contribution of highly excited states. If we knew the slope $a$ we
could find from Eq.~\eqref{14} the mass of ground state $m_0$
making use of the standard stability criterion on the
Borel~\cite{svz}. But we will prefer the opposite procedure:
Since $m_0$ is usually well known, it is interesting to calculate
numerical values of slopes for various radial trajectories.

\subsection{The scalar case}

The simplest scalar correlator is defined by replacing the vector
current in~\eqref{4} by the scalar one $j=\bar{q}q$. The OPE for
the scalar correlator reads~\cite{rry}
\begin{multline}
\label{15}
\Pi_s(q^2)=\frac{3}{8\pi^2}\left(1 + \frac{11}{3}\frac{\als}{\pi}\right) Q^2\ln\frac{Q^2}{\mu^2} + \frac{3}{Q^2}\mqq \\
+\frac{1}{8 Q^2}\gmv - \frac{22}{3}\frac{\pi\als}{Q^4}\qqb^2,
\end{multline}
where the coefficient in the last term is written in the large-$N_c$ limit.
We define the spectral representation as
\begin{equation}
\label{16}
\Pi_s(q^2)=\sum_{n}\frac{G^2 m_n^2}{q^2-m_n^2+i\varepsilon}.
\end{equation}
Here the constant $G^2$ represents a scalar analog of vector residue $F^2$.
Substituting the linear spectrum~\eqref{2} and repeating the operations of
the previous Section we will arrive at the following "combined" sum rule
in the scalar channel,
\begin{equation}
\label{17}
m_0^4 \lb e^{a/M^2} - 1\rb + 2 a m_0^2 = \L\lb a-m_0^2 + m_0^2\, e^{a/M^2}\rb-a^2\frac{e^{a/M^2} + 1}{e^{a/M^2} - 1},
\end{equation}
where
\begin{equation}
\label{18}
\L\equiv M^2\frac{2h_0 + \frac{h_3}{M^6}}{h_0 + \frac{h_2}{M^4} - \frac{h_3}{M^6}},
\end{equation}
and the coefficients $h_i$ are given in Table~1. The relation~\eqref{17}
represents a quadratic equation for the intercept $m_0^2$. The corresponding
two solutions are
\begin{equation}
\label{19}
m_0^2=\frac{a}{1-e^{a/M^2}}+\frac{\L}{2}\pm\frac{\sqrt{\L^2\lb e^{a/M^2} - 1\rb^2\!/4-a^2e^{a/M^2}}}{e^{a/M^2} - 1}.
\end{equation}
Below we will discuss both solutions.

\section{Numerical fits and predictions}

\subsection{Input parameters}

The numerical values of input parameters $h_i$ in Table~1 which we
will use are displayed in Table~2. Below these numbers are briefly
commented.

\begin{table}[ht]
\caption{\small The numerical values of coefficients $h_i$ in our fits.}
\begin{center}
$\begin{array}{|c|c|c|c|c|}
 \hline
 \text{Channel} & h_0 & h_1 & h_2 & h_3 \\
 \hline
 \rho & 1 & 0 & 0.032 & -0.030 \\
 \hline
 a_1 & 1 & -0.674 & 0.046 & 0.048 \\
 \hline
 f_0 & 1 & 0 & 0.032 & -0.095 \\
 \hline
\end{array}$
\end{center}
\end{table}

We set $h_0=1$ since taking the perturbative threshold
$s_0\rightarrow\infty$ (infinite number of radial states) we
should formally have  $\als\rightarrow0$ due to the asymptotic
freedom. This is tantamount to neglecting the loop corrections to
the unit operator in the OPE.

The values of gluon and quark condensates are taken from
Ref.~\cite{svz}: $\gmv=(330\,\text{MeV})^4,$ and
$\qqb=-(250\,\text{MeV})^3$. The first value is scale-independent
while the second one is taken roughly at the scale $\mu=1$~GeV.
From the Gell-Mann--Oakes --Renner relation,
$m_\pi^2f_\pi^2=-(m_u+m_d)\qqb$, with the pion mass
$m_\pi=140$~MeV~\cite{pdg} and pion weak decay constant
$f_\pi=92.4\,\text{MeV}$, one gets $m_u+m_d\approx
10.7\,\text{MeV}\Bigr|_{\mu=1\,\text{GeV}}$. We consider the
isospin limit for the masses of current quarks, $m_u=m_d\equiv
m_q$. Thus we get a numerical value for another renormalization
invariant condensate of dimension four, $\mqq=-(95.6\,\text{MeV})^4$.
All these inputs lead to the values of $h_1$ and $h_2$ in Table~2.

The operator $\als(\bar{q}q)$ has a small anomalous dimension. We
will regard the corresponding v.e.v. $\als\qqb$ as a constant. The
condensate $h_3$ in the $\rho$-channel is taken from
Ref.~\cite{svz}. All other $h_3$ can be then obtained from a
rescaling prescribed by the coefficients in the last column of
Table~1. Our numerical results, however, will be only slightly
dependent on $h_3$ or independent of it.

As was mentioned above we will regard $m_0$ as the mass of ground
state obtained within the classical SVZ sum rules. This reduces
the number of input parameters and makes the
method more attractive: Fixing the values of condensates in the
OPE, one can extract the values of both intercept $m_0$ and slope
$a$. The first step (extraction of $m_0$) is nothing but the
standard SVZ sum rule method while the second step represents our
extension of this method to the case of infinite linear radial spectrum.

\subsection{Vector mesons}

Our strategy is as follows. Consider the $\rho$-meson. We plot $m_0$ from
Eq.~\eqref{14} as a function of Borel parameter $M^2$ at different values
of $a$. A typical plot is presented in Fig.~1. These plots possess the
so-called "Borel window" --- a stability region near the minimum where $m_0$ is approximately
constant. We find the value of $a$ at which $m_0$ coincides with the value
of $m_\rho$ extracted in the usual SVZ sum rules~\cite{svz,rry}. The obtained
$a$ is our prediction for the slope. The slope of $\rho$-trajectory
turns out to be near $a_\rho=1.52\pm0.07$~GeV$^2$. The predicted masses
for this value of slope are presented in Table~3 together with a tentative
assignment to experimental data~\cite{pdg}. We give an uncertainty in mass related with
uncertainty in extraction of $m_\rho$ from the classical SVZ sum rules.
The latter uncertainty comes from uncertainty in values of vacuum condensates.
In order to avoid double counting of uncertainties we do not use
the uncertainties in condensates when calculate $a$. There are of course
uncertainties arising from the limits $N_c\rightarrow\infty$ and
$\als\rightarrow0$, and from assumed universality of residues. We estimate
these uncertainties at the level of 10\%.
\begin{figure}[ht]
    \includegraphics[scale=0.7]{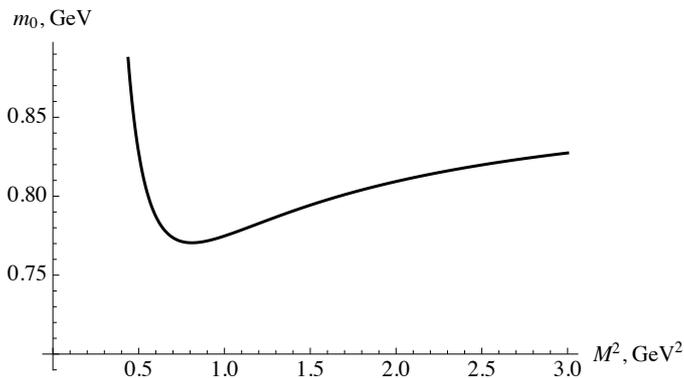}
    \caption{\small The mass of $\rho$-meson at $a=1.52\,\text{GeV}^2$ as a function of Borel parameter~\eqref{14}.}
\end{figure}

\begin{table}[ht]
\caption{\small The radial spectrum of $\rho$-mesons for the slope $a=1.52\pm0.07\,\text{GeV}^2$.
The masses are given in MeV.
The first 4 predicted states are tentatively assigned to the resonances $\rho(770)$, $\rho(1450)$,
$\rho(1900)$, and $\rho(2270)$~\cite{pdg} which presumably form the $S$-wave radial trajectory.}
\begin{center}
$\begin{array}{|c|c|c|c|c|c|}
 \hline
 n & 0 & 1 & 2 & 3 & 4\\
 \hline
 m_{\rho}\,\text{(th)} & 770\pm10 & 1450\pm20 & 1910\pm40 & 2230\pm50 & 2580\pm50 \\
 \hline
 m_{\rho}\,\text{(exp)} & 775 & 1465\pm25 & 1870\text{--}1920 & 2265\pm40 & \text{---} \\
 \hline
\end{array}$
\end{center}
\end{table}

In the axial-vector case, the stability region exists
only at large values of Borel parameter,
$M\rightarrow\infty$, see Fig.~2. The same situation takes place
within the classical SVZ sum rules~\cite{rry}. Normalizing our
$m_0$ to the value $m_{a_1}=1.15\pm0.04$~GeV obtained in
Ref.~\cite{rry}, we get $a_{a_1}=1.30\pm0.18$~GeV$^2$. The
corresponding mass spectrum is presented in Table~4.
\begin{figure}[ht]
    \includegraphics[scale=0.7]{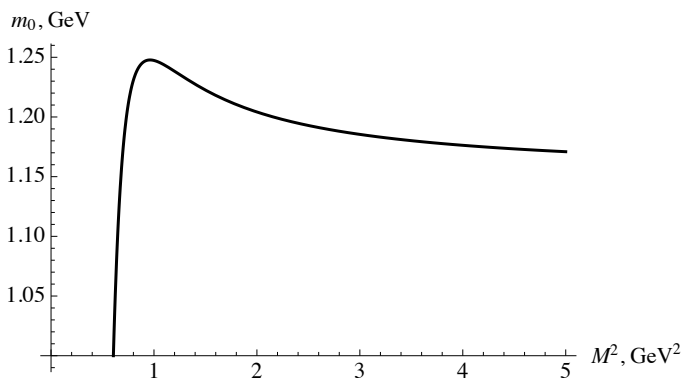}
    \caption{\small The mass of $a_1$ meson at $a=1.30\,\text{GeV}^2$.}
\end{figure}

Due to the second term in Eq.~\eqref{14}, however, an alternative
possibility appears. This additional contribution results in a
local maximum in Fig.~2. One can interpret the region near this
extremum as an "emergent" Borel window. The value of $m_0$ in this
region is surprisingly close to the mass of axial resonance
$a_1(1260)$~\cite{pdg} which is traditionally interpreted as an
axial partner of $\rho(770)$ if the chiral symmetry were not
spontaneously broken. Taking this solution as the ground state we
obtain an alternative prediction for the tower of radially excited
axial states. The corresponding spectrum is also shown in Table~4.

\begin{table}[ht]
\caption{\small The radial spectrum of $a_1$-mesons for the slope $a=1.30\pm0.18\,\text{GeV}^2$.
The first 4 predicted states are tentatively assigned to the resonances $a_1(1230)$, $a_1(1640)$,
$a_1(1930)$, and $a_1(2270)$~\cite{pdg}.}
%\centering
\begin{center}
$\begin{array}{|c|c|c|c|c|c|c|}
 \hline
 n & 0 & 1 & 2 & 3 & 4\\
 \hline
 m_{a_1}\,\text{(th)} & 1150\pm40 & 1620\pm60 & 1980\pm90 & 2280\pm120 & 2550\pm140 \\
 %\cline{2-6}
  & 1250 & 1690\pm50 & 2040\pm90 & 2330\pm120 & 2600\pm140 \\
 \hline
 m_{a_1}\,\text{(exp)} & 1230\pm40 & 1647\pm22 & 1930^{+30}_{-70} & 2270^{+55}_{-40} & \text{---} \\
 \hline
\end{array}$
\end{center}
\end{table}

\subsection{Scalar mesons}

The scalar case has two solutions~\eqref{19}. The first one (with plus sign)
exists at any $M^2$ while the second one appears above some positive value
of the Borel parameter, see Fig.~3. The first solution corresponds to the
value of scalar mass extracted in the classical SVZ sum rules~\cite{rry}.
The stability region lies at
$M\rightarrow\infty$ as in the axial case~\cite{sigma}. The standard SVZ
method gives $m_{f_0}=1.00\pm0.03$~GeV~\cite{rry}. Normalizing the first
solution to this prediction we obtain $a_{f_0}=1.38\pm0.07$~GeV$^2$.
If we substitute
this value of slope to the second solution we get mass of the "emerged"
lighter scalar meson, $m_{f_0}\approx0.62$~GeV. Our solution predicts thus two
parallel scalar trajectories. The ground state on the first trajectory
can be identified with $f_0(980)$ and on the second one with $f_0(500)$~\cite{pdg}.
The existence of two parallel radial scalar trajectories seems to agree
with the experimental data~\cite{phen}. The masses of predicted radial
states and a tentative comparison with the observed scalar mesons for two
trajectories are displayed in Tables~5 and~6, correspondingly.
\begin{figure}[ht]
    \includegraphics[scale=0.7]{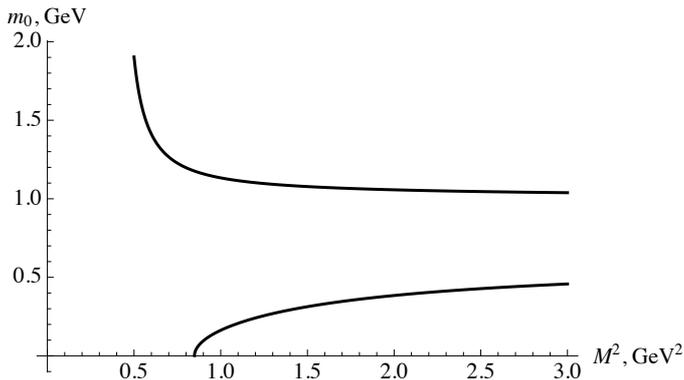}
    \caption{\small The mass of scalar meson at $a=1.38\,\text{GeV}^2$.}
\end{figure}

\begin{table}[ht]
\caption{\small The radial spectrum of the first $f_0$-trajectory for the slope $a=1.38\pm0.07\,\text{GeV}^2$.
The first 5 predicted states are tentatively assigned to the resonances $f_0(980)$, $f_0(1500)$,
$f_0(2020)$, $f_0(2200)$, and $X(2540)$~\cite{pdg}.}
\begin{center}
$\begin{array}{|c|c|c|c|c|c|}
 \hline
 n & 0 & 1 & 2 & 3 & 4\\
 \hline
 m_{f_0}\,\text{(th 1)} & 1000\pm30 & 1540\pm20 & 1940\pm40 & 2270\pm50 & 2560\pm50 \\
 \hline
 m_{f_0}\,\text{(exp 1)} & 990 \pm 20 & 1504 \pm 6 & 1992 \pm 16 & 2189 \pm 13 & 2539 \pm 14^{+38}_{-14} \\
 \hline
\end{array}$
\end{center}
\end{table}

\begin{table}[ht]
\caption{\small The radial spectrum of the second $f_0$-trajectory for the slope $a=1.38\pm0.07\,\text{GeV}^2$.
The first 5 predicted states are tentatively assigned to the resonances $f_0(500)$, $f_0(1370)$,
$f_0(1710)$, $f_0(2100)$, and $f_0(2330)$~\cite{pdg}.}
\begin{center}
$\begin{array}{|c|c|c|c|c|c|}
 \hline
 n & 0 & 1 & 2 & 3 & 4\\
 \hline
 m_{f_0}\,\text{(th 2)} & 620 & 1330\pm30 & 1780\pm40 & 2130\pm50 & 2430\pm60 \\
 \hline
 m_{f_0}\,\text{(exp 2)} & 400\text{--}550 & 1200\text{--}1500 & 1723^{+6}_{-5} & 2101 \pm 7 & 2300\text{--}2350 \\
 \hline
\end{array}$
\end{center}
\end{table}

\section{Discussions and conclusions}

We have put forward a new extension of borelized SVZ sum rules.
Using an example of a simple model with minimum of inputs, we demonstrated
how this extension allows to extract the value of slope of linear
radial trajectories from static characteristics of QCD vacuum ---
the vacuum condensates --- in the large-$N_c$ limit of QCD.
The obtained slopes for the light non-strange vector, axial,
and scalar trajectories agree well with the phenomenology.
This may justify {\it a posteriori} the approximations and assumptions made
for a simple demonstration of the method.
Our analysis confirms a known hypothesis that the slope of radial
trajectories is approximately universal for all light non-strange
mesons. We thus obtained an independent estimate for its
value, $a=1.4\pm0.1\,\text{GeV}^2$. This value is consistent with a typical
phenomenological estimate, $a=1.25\pm0.15\,\text{GeV}^2$~\cite{phen}.

In the limit of infinite Borel parameter, our borelized sum rule
for the vector and axial channel becomes one of the usual planar sum rules
which were considered many times in the past~\cite{sr}. Indeed, taking
the limit $M^2\rightarrow\infty$ in Eq.~\eqref{14} we get
\begin{equation}
\label{20}
m_0^2=-h_1+\frac{a}{2}.
\end{equation}
The sum rule~\eqref{20} represents a planar analog of the first Weinberg sum rule.
For comparison, we display in Table~7 some fits based on relation~\eqref{20}.

\begin{table}[ht]
\caption{\small The radial spectrum of vector and axial mesons in the limit $M^2\to\infty$.}
\begin{center}
$\begin{array}{|c|c|c|c|c|c|c|c|}
 \hline
 \text{Meson} & h_1,\,\text{GeV}^2 & m_0 & a,\,\text{GeV}^2 & m_1 & m_2 & m_3 & m_4 \\
 \hline
 \rho & 0 & 770 & 1.19 & 1330 & 1720 & 2040 & 2310 \\
 \hline
 a_1  & -0.674 & 1150 & 1.30 & 1620 & 1980 & 2280 & 2550 \\
 %\cline{3-8}
 & & 1250 & 1.78 & 1830 & 2260 & 2630 & 2940 \\
 \hline
\end{array}$
\end{center}
\end{table}

In the scalar channnel, our borelized sum rule has no analog in the
planar sum rules without borelization. The reason is that taking the
limit $M^2\to\infty$ in Eq.~\eqref{17} we arrive at identity $0=0$.
Applying this limit to the solutions~\eqref{19} we get
\begin{equation}
\label{21}
m_0^2=\frac{a}{2}\pm\frac{1}{2}\sqrt{\frac{a^2}{3}-8 h_2}.
\end{equation}
The relation~\eqref{21} gives an analytical expression for the masses
of two lightest scalar states which we obtained from Fig.~3. It is seen
that these masses do not depend on the condensate $h_3$.

The prediction of the second scalar trajectory is a rather
surprising feature of our borelized planar sum rules. The ground
state on the second radial trajectory turns out to be significantly
lighter than on the first trajectory. It looks tempting to identify
this state with the elusive $\sigma$ (called also $f_0(500)$) meson~\cite{pdg}.
The lightest scalar state in the standard SVZ sum rules lies near 1~GeV~\cite{rry}
and cannot be made significantly lighter within this method~\cite{sigma}.
Our extension of the SVZ method leads thus to a new result. It is interesting to
check whether a similar result appears in the framework of unborelized
planar sum rules. A recent analysis of Ref.~\cite{AS} gives a positive answer.
A light scalar state near 0.5~GeV, however, emerged in Ref.~\cite{AS} in a
different way.

\end{document}